\documentstyle[epsf,twocolumn,prl,aps]{revtex}
\begin{document}
\draft
\title{Zeeman Coupling and Abnormal Thermal Conductivities\\
in BSCCO Superconductors}
\author{Qiang-Hua Wang$^{1,2}$, and Z. D. Wang$^{2}$}
\address{$^{1}$Department of Physics and National Laboratory of Solid State
Microstructures,\\
Nanjing University, Nanjing 210093, China\\
$^{2}$Department of Physics, University of Hong Kong, Pokfulam, Hong Kong,
China}
\date{September 27, 1999}

\twocolumn[\hsize\textwidth\columnwidth\hsize\csname@twocolumnfalse\endcsname
\maketitle

\begin{abstract}
Using the path-integral formulation we derive microscopically a
Ginzburg-Landau free energy with a Zeeman coupling between the magnetic
field and the orbital angular momentum of the Cooper pairs in a
superconductor with singlet pairing in the $d_{x^{2}-y^{2}}$- and the
sub-dominant $d_{xy}$- channels. The Zeeman coupling induces a
time-reversal-symmetry-breaking pairing state.
Based on careful examinations of the energy
gain due to the Zeeman coupling, the energy lost due
to the kinetic energy of the excess superfluid, and the Doppler energy shift
for quasi-particle excitations,
we present a coherent interpretation for the puzzling and
conflicting thermal conductivites observed at above $5K$ (K. Krishana,
{\it et al}, Science {\bf 277}, 83(1997)) and
at sub-Kelvins (H. Aubin, {\it et al}, Phys. Rev.
Lett. {\bf 82}, 624 (1999)) in BSCCO superconductors.
\end{abstract}

] %
\vskip2pc \narrowtext

Recently, the anomalous low-temperature thermal conductivity versus the
magnetic field observed in BSCCO superconductors \cite
{Krishana,Aubin} has been stimulating considerable interest in the study of
the pairing states in the cuprates \cite{Laughlin,Kubert,Wang}. Although it is
widely accepted that the pairing is singlet, and that the dominant pairing
channel is the $d_{x^{2}-y^{2}}$-channel \cite{Tsui}, it is still
interesting to investigate whether there is a bulk time-reversal-symmetry ($%
\top $) breaking pairing states involving the $d_{x^{2}-y^{2}}$-channel and
a sub-dominant channel, such as the $s$- or $d_{xy}$-channel \cite{T-breaking}.
The abnormal field dependence of the low temperature (but above $5$K) thermal
conductivity $\kappa _{e}$ in BSCCO \cite{Krishana} is remarkable, in that $%
\kappa _{e}$ is a bulk quantity, and the observed plateau in its field
dependence suggests a possible bulk $\top $-breaking states
\cite{Laughlin,Wang}, so that a full gap is opened at the Fermi surface for
quasi-particle excitations. On the other hand, at even lower (sub-Kelvin)
temperatures, another anomaly arises: Instead of decreasing with increasing
magnetic field $B$, $\kappa _{e}\propto \sqrt{B}$ \cite{Aubin}. This
behavior implies a pure $d_{x^{2}-y^{2}}$-wave pairing state, with
which quasiparticle states are popularized at the Fermi surface along the
nodal direction by the supercurrent around the vortices, with the induced
density of states $\propto \sqrt{B}$ \cite{Volovik}. Thus these
two sets of $\kappa_e$ data seemingly point towards conflicting pairing states,
and for a long time, remain puzzling.

It is now clear in this Letter that a Zeeman coupling between the
magnetic field and the internal motion of Cooper pairs may shed remarkable
lights on the abnormal thermal conductivities \cite{Note1}.
Physically, the Zeeman coupling is
expected based on general footing \cite{Sigrist}, in that a Cooper pair carrying an
internal angular momentum also carries an internal magnetic moment, which is
coupled to the magnetic field. This effect may be expected to cancel
out after averaging over the Fermi surface, but will be
examined seriously in this work. The main findings in this Letter are: (i) A
sound microscopic derivation of the Zeeman coupling term in the GL theory;
(ii) With the Zeeman coupling and the Doppler energy shift due to
the superfluid for quasi-particle excitations,
a coherent interpretation for both sets of the $\kappa _{e}$
data mentioned above is presented. The seemingly conflicting pictures
regarding the pairing states relevant to $\kappa_e$ are unified.

For definiteness, we consider a superconductor with pairing in the $%
d_{x^{2}-y^{2}}$- and $d_{xy}$-channels. The pairing function is assumed to
be $\Delta _{{\bf kq}}=D_{{\bf q}}\cos 2\theta _{{\bf k}}+D_{{\bf q}%
}^{\prime }\sin 2\theta _{{\bf k}},$where ${\bf k}$ and ${\bf q}$ describe
the internal motion and center-of-mass motion of the Cooper pair, and $D_{%
{\bf q}}$ and $D_{{\bf q}}^{\prime }$ are the ${\bf q}$-modes of the order
parameters in the $d_{x^{2}-y^{2}}$- and $d_{xy}$-channels, respectively.
Here $\theta _{{\bf k}}$ is the angle between ${\bf k}$ and, say, the $a$%
-axis of the $ab$-plane. As is well accepted, the singlet pairing with a
dominant $d_{x^{2}-y^{2}}$-channel is present in high temperature
superconductors. The pairing interaction responsible for the pairing
function is assumed to be $V_{{\bf k,k^{\prime }}}\equiv V_{D}\cos 2\theta _{%
{\bf k}}\cos 2\theta _{{\bf k^{\prime }}}+V_{D^{\prime }}\sin 2\theta _{{\bf %
k}}\sin 2\theta _{{\bf k^{\prime }}}.$ The Bardeen-Cooper-Shrieffer (BCS)
effective Hamiltonian reads, 
\begin{eqnarray}
H &=&\int_{{\bf x}}\Psi _{\sigma }^{\dagger }({\bf x})h_{0}({\bf x})\Psi
_{\sigma }({\bf x})  \nonumber \\
&&-\int_{{\bf x,x^{\prime }}}\Psi _{\uparrow }^{\dagger }({\bf x})\Psi
_{\downarrow }^{\dagger }({\bf x^{\prime }})V({\bf x-x^{\prime }})\Psi
_{\downarrow }({\bf x^{\prime }})\Psi _{\uparrow }({\bf x}),
\end{eqnarray}
where $h_{0}({\bf x})=(-i\nabla -e{\bf A})^{2}/2m$ is the single-particle
Hamiltonian (with $\hbar=c=1$)
and $V$ is the pairing interaction (in real space). Repeated
indices imply summation. The other notations are standard. Using the
coherent Fermion path integral formulation of the BCS theory, the
Ginzburg-Landau free energy can be obtained by a loop expansion at the
saddle-point of the effective action after performing the standard
Hubbard-Stratonavich transform that decouples the pairing interaction. It
can be written as, up to the forth-order in the order parameters and
symbolically, 
\begin{eqnarray}
F &\approx &\int \Delta ^{\ast }V^{-1}\Delta -T{\rm Tr}(g\Delta g^{\ast
}\Delta ^{\ast })  \nonumber \\
&&+T{\rm Tr}(g\Delta g^{\ast }\Delta ^{\ast }g\Delta g^{\ast }\Delta ^{\ast
}),  \label{Eq:loop}
\end{eqnarray}
where $g$ is the normal state single-particle Matsubara Green's function and 
$V^{-1}$ denotes the inverse operator of $V$. In detail, 
\begin{equation}
g({\bf x,x^{\prime }};i\omega _{n})\approx g_{0}({\bf x,x}^{\prime };i\omega
_{n})\exp [i{\bf A(x}^{\prime }{\bf )\cdot (x-x}^{\prime }{\bf )}]
\label{Eq:g}
\end{equation}
where $g_{0}({\bf x,x}^{\prime };i\omega _{n})$ is the Green's function at $%
A=0$, and $\omega _{n}=(2n+1)\pi T$. The usual semi-classical approximation
is used. The second term in Eq.(\ref{Eq:loop}) is, after some manipulations, 
\begin{eqnarray*}
F^{(2)} &=&-T\sum_{{\bf k,k}^{\prime }}\int_{{\bf R,r}}[g_{0}(-{\bf k}%
^{\prime }+{\bf \Pi }^{\ast }/2-e\delta {\bf A};i\omega _{n})\Delta
_{{}}^{\ast }({\bf k}^{\prime };{\bf R}) \\
&&{\rm e}^{i{\bf k}^{\prime }{\bf \cdot r}}]\times \lbrack g_{0}^{\ast }(-%
{\bf k+\Pi }/2+e\delta {\bf A};i\omega _{n})\Delta ({\bf k};{\bf R}){\rm e}%
^{i{\bf k\cdot r}}]
\end{eqnarray*}
where ${\bf \Pi }=-i{\bf \nabla }_{{\bf R}}-2e{\bf A(R)}$ is the gauge
invariant gradient, $\Delta ({\bf k};{\bf R})=\Delta _{{\bf kq}}{\rm e}^{i%
{\bf q\cdot R}},$ and $\delta {\bf A(r)}={\bf r\cdot \nabla }_{{\bf R}}{\bf %
A(R})/2\approx {\bf B}\times {\bf r/}4$ is a correction to the naive
semi-classical approximation that is usually neglected. Here we have assumed 
${\bf A}=(-By/2,Bx/2,0)$ in the calculation of $\delta {\bf A(r}).$ This
term generates the Zeeman coupling we broadcasted earlier. So let us first
look into the energy in the first order of $\delta {\bf A(r}).$ After some
algebra, we find it is, 
\begin{equation}
F_{{\rm z}}=T\int_{{\bf R},{\bf r}}[\psi _{n}^{\ast }({\bf r;R})e{\bf B\cdot
L(r)}\phi _{n}({\bf r;R})+{\rm c.c}]/4,
\end{equation}
where ${\bf L(r)=r\times \nabla }_{{\bf r}}/i$ is the angular momentum
operator for the internal motion of the Cooper pair, and $\psi _{n}$ and $%
\phi _{n}$ can be imagined as the dressed wave function for the relative
motion in a Cooper pair (propagating in the imaginary time with a frequency $%
\omega _{n}$), defined by $\psi _{n}({\bf r;R})\approx \alpha _{n}\tilde{%
\Delta}({\bf r;R}),\phi _{n}({\bf r;R})\approx \beta _{n}\tilde{\Delta}({\bf %
r;R}),$with 
\begin{eqnarray}
\alpha _{n} &=&-\int \frac{N(\epsilon )d\epsilon }{(i\omega _{n}+\epsilon )}%
;\ \ \ \beta _{n}=\int \frac{N(\epsilon )d\epsilon }{(i\omega _{n}+\epsilon
)^{2}}; \\
\tilde{\Delta}({\bf r;R}) &\approx &-J_{2}(k_{F}r)[D({\bf R})\cos 2\theta _{%
{\bf r}}+D^{\prime }({\bf R})\sin 2\theta _{{\bf r}}],  \label{Eq:DressedD}
\end{eqnarray}
where $J_{n}$ is the Bessel function, $N(\epsilon)$ is the normal state density of
states, and $\theta _{{\bf r}}$ is the direction angle of ${\bf r.}$ After
some further manipulations, we find 
\begin{equation}
F_{z}=\int_{{\bf R}}-iN(0)(B/B_{\ast })(D^{\ast }D^{\prime }-{\rm c.c.}),
\label{Eq:Fz}
\end{equation}
with $B_{\ast }$ a characteristic magnetic field given by 
\begin{equation}
B_{\ast }^{-1}\sim (m_{F}/\pi )(N(0)/\rho
_{e})z(T)\int_{0}^{k_{F}r_{0}}[J_{2}(x)]^{2}xdx,  \label{Eq:Bstar}
\end{equation}
\begin{equation}
z(T)=\int_{\epsilon _{1,}\epsilon _{2}}\frac{N(\epsilon _{1})dN(\epsilon
_{2})/d\epsilon _{2}}{2N(0)^{2}}\frac{f(\epsilon _{1})+f(\epsilon _{2})-1}{%
\epsilon _{1}+\epsilon _{2}}.  \label{Eq:z}
\end{equation}
Here $m_{F}=ev_{F}\lambda _{F}/2$ is a characteristic magnetic moment$,$ $%
\omega _{c}$ is the BCS energy cutoff, $\rho _{e}$ is the density of charge
carriers and $r_{0}\sim 2\pi v_{F}/\omega _{c}$ is a length cutoff because
of the BCS truncation ($\delta k\sim \omega _{c}/v_{F}\sim 2\pi /\delta r$).
Here $v_{F}$, $\lambda _{F}$ and $f(\varepsilon )$ are the Fermi velocity,
length and distribution, respectively. Although $z(T)$ could not be
evaluated exactly, an upper bound exists, 
\[
|z(T)|<(W/8T)\int_{\epsilon }|dN(\epsilon )/d\epsilon |/N(0), 
\]
where $W$ is of the order of the band width. The fact that $|z(T)|$ scales
at most as $T^{-1}$is important to extrapolate it to low temperatures, as
compared to the kinetic terms to be discussed below. Clearly, $z(T)=0$ if $%
N(\epsilon )$ is an even function of $\epsilon $. Thus for a nonzero Zeeman
coupling as described by Eq.(\ref{Eq:Fz}), a necessary condition is the
particle-hole asymmetry in the density of states \cite{Koyama,Note:asymmetry}.
Such asymmetry might be related to strong coupling effects, although our
derivation is in the weak coupling limit.

The other contributions to the free energy are usual \cite{Wang,NaiveTheories}.
For our purpose, we present explicitly the kinetic energy of the superfluid, 
\begin{equation}
F_{k}\approx \int_{{\bf R}}N(0)(v_{F}^{2}/16)\Gamma ({|{\bf \Pi }D|^{2}+|%
{\bf \Pi }D^{\prime }|^{2}),}  \label{Eq:Fu}
\end{equation}
where $\Gamma =7\zeta (3)/(8\pi ^{2}T^{2})$ with $\zeta $ being the Riemann
zeta function. It is easy to see that the dimensionless factor $\delta
k=8\Phi _{0}/(\pi v_{F}^{2}B_{\ast }\Gamma )$ is a measure of the relative
importance of the Zeeman energy as compared to the kinetic energy, recalling
that $F_{z}$ can also be casted into a similar form to that of $F_{k}$ ({\it %
i.e}., in terms of gradients) because of the identity $[\Pi _{x},\Pi
_{y}]=2ieB$. Strictly speaking, the parameters in $F_{z}$ and $F_{k}$ are
defined near $T_{c}$ ($=T_{D}$ here). But if extrapolated to low
temperatures, we may expect that $|\delta k|=T/T_z$ with an unknown constant $T_z$
(applicable at $T\lesssim T_z$).

The complete GL free energy can be casted in the following form \cite{Wang},
\begin{eqnarray}
F &\approx &E_{c}\int_{{\bf r}}-|d|^{2}-\alpha _{d^{\prime }}|d^{\prime
}|^{2}+|{\bf \pi }d|^{2}+|{\bf \pi }d^{\prime }|^{2}  \nonumber \\
&&-i\delta kb(d^{\ast }d^{\prime }-{\rm c.c.})+{\bf |}d|^{4}/2+{\bf |}%
d^{\prime }|^{4}/2  \nonumber \\
&&+|d|^{2}|d^{\prime }|^{2}/3+(d^{\ast }d^{\prime }+{\rm c.c.})^{2}/6+\kappa
^{2}b^{2},  \label{Eq:F}
\end{eqnarray}
where $E_{c}=H_{c}^{2}\xi ^{2}/4\pi $ with $H_{c}$ and $\xi $ being the
thermodynamic critical field and the coherence length, respectively, when
the $d_{xy}$-channel is absent. All quantities under the integration symbol
are now dimensionless: $\alpha _{d^{\prime }}=\ln (T_{D^{\prime }}/T)/\ln
(T_{D}/T)$ with $T_{i}=(2\omega _{c}e^{\gamma }/\pi )e^{-2/N(0)V_{i}}$ being
the bare critical temperature for the $i$-th order parameter ($i=D,D^{\prime
}$, and $\gamma $ is the Euler constant); $d=D/D_{0}$ and $d^{\prime
}=D^{\prime }/D_{0}$ are the normalized order parameters, with $D_{0}$ being
the value of $D$ at zero magnetic field and in the absence of $D^{\prime }$, 
${\bf \pi }=\xi {\bf \Pi }={\bf -i\nabla -a}$ is the dimensionless gauge
invariant gradient, $\kappa $ is the GL parameter, ${\bf b=\nabla \times a=B/%
}B_{0}$ is the dimensionless magnetic field with $B_{0}=\Phi _{0}/2\pi \xi
^{2},$ and finally $r=R/\xi$. The merit of the dimensionless form is to
hide all irrelevant parameters, but we shall also use dimensionless and
dimensioned forms inter-changeably.

The Zeeman term violates both parity as well as time-reversal symmetry
\cite{Laughlin}, leading to many nontrivial consequences. First, in the presence
of a magnetic field, there will be a $\top $-breaking $d_{x^{2}-y^{2}}\pm
id_{xy}$-pairing state, with a minimum gap for quasi-particle excitations at
the Fermi surface given by $\min (|D|,|D^{\prime }|)$ \cite{Wang}.
Second, because the vector potential is coupled to the supercurrent, the
Zeeman coupling should induce a spontaneous edge supercurrent \cite{Laughlin}.
This effect may be used for an experimental verification of the Zeeman
coupling. Third, there will also be a bulk first-order transition from the
Meissner state to the mixed state with a {\it finite} density of vortices if 
$\alpha _{d^{\prime }}=1$, as discussed earlier \cite{pwave}.

Let us now discuss the relevance of the Zeeman coupling in the abnormal
thermal conductivity observed in the cuprates. To be consistent with
experiments, $T_{D}\gg T_{D^{\prime }}$. At low temperatures, $D_{0}\sim
2.13T_{D}$ is essentially independent of temperature for $d_{x^{2}-y^{2}}$
-wave superconductors. From Eq.(\ref{Eq:F}), there would be a
zero-field $\top $-breaking pairing transition at $\alpha _{d^{\prime
}}=1/3$, or at a temperature $T_{\ast }=\sqrt{T_{D^{\prime }}^{3}/T_{D}}$
if the above relation of $\alpha _{d^{\prime }}$ were valid at all
temperatures. Such a zero-field $\top $%
-breaking pairing state posed the major difficulty in a previous theoretical
study \cite{Wang} to explain the sub-Kelvin $\kappa _{e}$ data \cite{Aubin}.
In fact, the derived temperature dependence of the parameters
in the GL theory is restricted near the
superconducting transition temperature, and the above $\alpha_{d'}$
is invalid at low temperatures.
Instead, in the spirit of the two-fluid model for a general superconductor
\cite{Tinkham},
the temperature dependence of $\alpha _{d^{\prime }}$ is better
replaced by $\alpha _{d^{\prime }}\propto (1-T^2/T_{D'}^2)/(1-T^2/T_D^2)$.
Since the $\top$-breaking transition (in the bulk) has not been observed
experimentally yet, we demand that
$T_{\ast }=0$, leading to $\alpha_{d'}=(1-T^2/T_{D'}^2)/3$.
In this realistic case, $d^{\prime }$ can only
be induced by the Zeeman coupling (and is always out of phase with $d$). This
permits a perturbative treatment of $d^{\prime }$. At low fields and in the
London limit, $|d|\sim 1$ at $1\ll r\ll \kappa $ where $r$ is the distance
off a vortex core, while the field $b $ is essentially uniform. Let us set
$d^{\prime }=-i\eta d\ {\rm sgn}(\delta k)$ (even in the
vortex state), and find $\eta $ variationally, which is
at least a qualitative estimation.
The dimensionless excess energy density due
to the induced $d^{\prime }$ is estimated as, to the second order in
$\eta$ : 
\begin{equation}
\delta f=[1/3-\alpha _{d^{\prime }}+b\ln (1/b)]\eta ^{2}-2|\delta k|b\eta ,
\end{equation}
where the core energy of vortices is neglected, and the $b\ln (1/b)$ term is
the kinetic energy of $d^{\prime }$. The latter is obtained as follows. In
the absence of $d^{\prime }$, we have a well-known magnetization curve at
low and intermediate fields, $H=B+(H_{c1}/\ln \kappa )\ln (B_{0}/B)$ where
$H_{c1}$ is the lower critical field \cite{Tinkham}. The kinetic energy
density of $d$ due to its superfluid is roughly given by $B(H-B)/4\pi $ from
the Virial theorem. The kinetic energy density of $d^{\prime
} $ is $\eta ^{2}$ times that of $d$, which enters our $\delta f$ as it
stands after proper normalization. Thus the optimum $\eta $ is given by 
\begin{equation}
\eta =\frac{3|\delta k|b}{1-3\alpha _{d^{\prime }}+3b\ln (1/b)}=\frac{%
3(T/T_{z})b}{T^{2}/T_{D^{\prime }}^{2}+3b\ln (1/b)}.  \label{Eq:Result}
\end{equation}
There would be an induced full gap
at the Fermi surface of the size $\eta D_0$
{\it in the bulk} (excluding the vortex cores).
However, as found earlier \cite{Aubin,Kubert},
the Doppler energy shift $E_{\rm Doppler}$ turns out to be essential
to explain the low-field sub-Kelvin $\kappa_e$.
Roughly speaking,
$E_{{\rm Doppler}}\sim P_{F}v_{s}\sim (a/\sqrt{\ln\kappa})D_{0}\sqrt{b\ln(1/b)}$,
where $P_{F}$ is the Fermi momentum, $v_{s}\sim \hbar/2mR_v$
is the characteristic superfluid velocity with
an inter-vortex spacing $R_{v}$, and $a$
is a vortex-lattice dependent constant of order unity.
Here we have included the logarithmic correction in $b$, in the same spirit
as for the superfluid kinetic energy, in order to take into account
the suppression of the superfluid arising from surrounding vortices.
In a nodal $d_{x^{2}-y^{2}}$-wave pairing state,
$E_{\rm Doppler}$ is responsible for the
vortex induced density of states at the Fermi surface, scaling as
$\sqrt{B}$ at low fields \cite{Volovik}.
Collecting both effects, the net gap at the Fermi surface
is $\Delta_{\min}=\eta D_0-E_{\rm Doppler}$. Of course, $\Delta_{\min}$
blocks (promotes) quasi-particle excitations if it is positive (negative).
$\kappa_e(B,T)/\kappa_e(0,T)\propto (1/T)\exp(-\Delta_{\min}/T)$ would develop a kink
at $\Delta_{\min}\sim T$. In the absence of the Doppler effect,
the kink field $B_k$ is given by 
$T=T_{D'}\sqrt{3(B_k/B_0)[D_0/T_z-\ln(B_0/B_k)]}$. At
$B_k\gg B_{\times}=B_0{\rm e}^{-D_0/T_z}$,
we would have the celebrated power law
$B_k\propto T^2$ \cite{Krishana,Wang}.
At $T\rightarrow 0$, $\eta D_0\propto T$ due to the emerging
role of the kinetic energy in $\eta$,
and the power law is violated.
The Doppler effect emerges at low fields, with $E_{\rm Doppler}\propto
\sqrt{b}$ within logarithmic accuracy, which dominates over $\eta D_0$,
and one would expect $\kappa_e\propto \sqrt{B}$.
However, the competition is complicated by the role of
temperature in $\eta$, and the value of $a/\sqrt{\ln\kappa}$ in
$E_{\rm Doppler}$. In order to proceed, we take
reasonable parameters: $T_D=100K$, $T_{D'}=8K$, $T_z=10K$
and $a/\sqrt{\ln\kappa}=0.1$.
The field dependence of $\Delta_{\min}$ is shown in Fig.1(a) at
the specified temperatures, where $\Delta_{\min}>20K$ is disposed,
as we are treating $d'$ perturbatively.
$\Delta_{\min}$ rises quasi-linearly at above $2K$,
but $\Delta_{\min}<0$ {\it for good} at $T=0.1K$.
This marks the Zeeman regime
at above $2K$, the Doppler regime at sub-Kelvins,
and the cross-over regime (grey zone) at intermediate 
temperatures, as highlighted in Fig.1(b).
In the Zeeman regime, as we already declared,
a kink in $\kappa_e(B)$ develops at $\Delta_{min}\sim T$. The kink points
extracted from Fig.1(a) is plotted in Fig.1(b) (squares), where the
dotted line is a fit to the power law $B_k\propto T^2$.
In the Doppler regime, we naturally expect an increasing $\kappa_e(B)$,
since $\kappa_e\propto -N(0)\Delta_{\min}$ (with $\Delta_{\min}<0$).\nolinebreak
Moreover, the scaling law
$\kappa_e\propto \sqrt{B}$ is already visible at $T=0.1K$ in Fig.1(a),
{\it as if in a pure $d_{x^2-y^2}$-wave state}.
\begin{figure}
\epsfxsize=8cm
\epsfbox{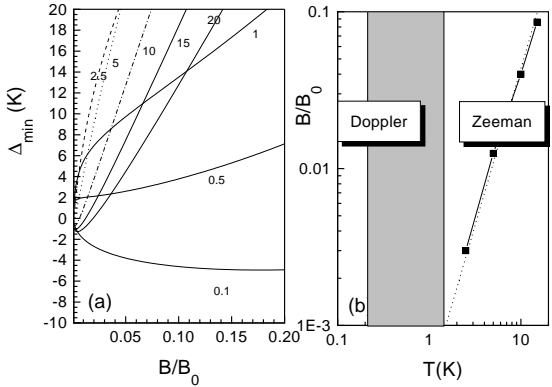}
\caption{
(a) The field dependence of $\Delta_{\min}$. The numbers represent the temperatures
(in unit of K); (b) The temperature dependence of $B_k$ (squares) extracted from
(a). See the text for details.}
\end{figure}
On the other hand, the impurity scattering
was indicated to be important {\it in a pure} $d_{x^2-y^2}$ superconductor
in the Doppler regime to
explain a monotonically increasing $\kappa_e(B)$ at low temperatures, and a
nonmonotonic field dependence at higher temperatures \cite{Aubin,Kubert},
but without a $\top$-breaking
state at higher temperatures and fields, nothing could be said for
the power law in the kink field. 
Therefore, the present theory nicely bridges the `gap' between
the the existing conflicting pictures regarding the
pairing states \cite{Aubin,Laughlin,Kubert,Wang}.

Finally, we remark that it was argued \cite{Franz} that the
competition between the quasi-particle density of states
and the vortex scattering effect
could be responsible for the plateau in $\kappa_e(B)$.
However, this scenario alone is difficult to account for the celebrated
power law $B_k\propto T^2$ \cite{Krishana}. We think that
including the vortex scattering effects in our theory would render even more
promising agreement with the experiments. On the other hand,
the role of a sub-dominant $s$-wave pairing can be ruled out,
as it does not participate the Zeeman coupling by symmetry, so that it
can only be induced locally by inhomogeneities \cite{Wang}.
In contrast, the Zeeman coupling is effective in the {\it bulk} because $B$ is
essentially uniform in the high $\kappa $ limit.

This work was supported by National Natural Science Foundation of China and
the RGC grant of Hong Kong under No. HKU7116/98P and HKU 7144/99P. We also
thank Dr. Shun-Qing Shen for helpful discussions.


\begin{references}
\bibitem{Note:asymmetry}  This same condition was discussed
in Ref.\cite{Koyama} but we need no
`ghost' $p$-wave pairing interaction (that does not lead to a $p$-wave gap).
\end{references}
\end{document}